# Kinetics of drug release from clay using enhanced sampling methods

Ana Borrego-Sánchez[a], Jayashrita Debnath[a], Michele Parrinello[a,*]

[a] Center for Human Technologies, Italian Institute of Technology (IIT)
   Via Enrico Melen 83, 16152 Genoa (Italy)

*Corresponding author.
E-mail: michele.parrinello@iit.it

## Abstract

A key step in the development of a new drug is the design of drug-excipient complexes that lead to an optimal drug release kinetics. Computational chemistry, specifically enhanced sampling molecular dynamics methods, can play a key role in this context by minimizing the need for expensive experiments and reducing cost and time. Here, we show that recent advances in enhanced sampling methodologies can be brought to fruition in this area. We demonstrate the potential of these methodologies by simulating the release kinetics of the complex praziquantel-montmorillonite in water. Praziquantel finds promising applications in the treatment of schistosomiasis, but its biopharmaceutical profile needs to be improved and a cheap material like the montmorillonite clay would be a very convenient excipient. We simulate the drug release both from surface and interlayer space and find that the time it takes to the drug to diffuse from the interlayer space to the solution controls the release time.

## Graphical abstract

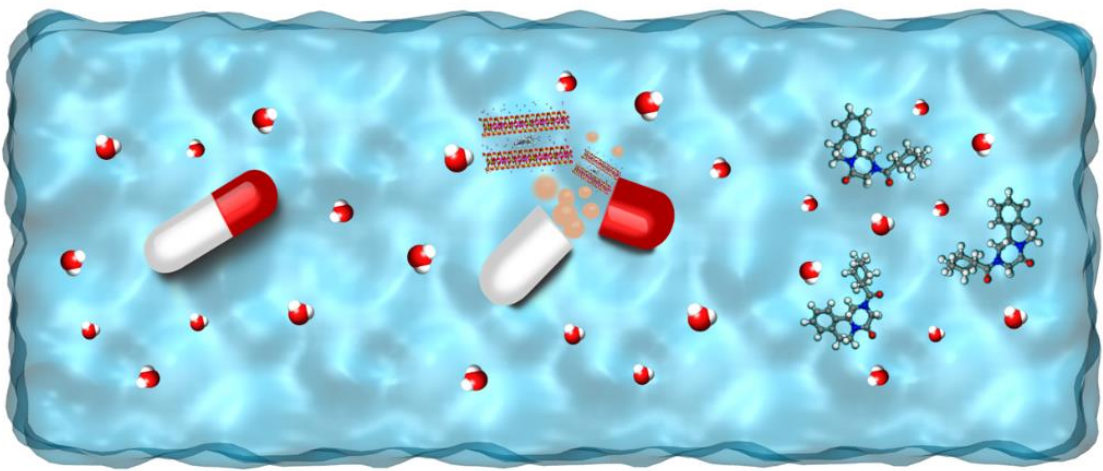

# 1. Introduction

One of the challenges in pharmacological studies is the design of new drug delivery systems that have a release profile able to reduce doses and minimize side effects. This implies developing new drug-excipient combinations. An example of current relevance is praziquantel. This is the drug of choice in the treatment of schistosomiasis, which is a widely spread but neglected tropical disease. Worldwide more than 700 million people are exposed to the parasite that carries this disease and 240 million are infected, mainly in the tropical and subtropical regions of developing countries (WHO, 2021). Praziquantel is administered orally, but due to its low solubility, high doses are needed to obtain effective concentrations in the blood. This causes side effects and leads to drug resistance (Cioli and Pica-Mattoccia, 2003). To overcome these limitations, scientific research is needed to improve the praziquantel aqueous solubility possibly by using cheap excipients to keep the cost of the medicine low, which is a major issue in developing countries. In this sense, excipients based on clay minerals are attractive candidates. In fact, montmorillonite has been already used in the pharmaceutical practice since it is abundant and has many advantageous properties. It is cheap, safe, non-toxic, biocompatible, and highly adsorbent, and it can encapsulate the drug in the nanosized interlayer spaces (Aguzzi et al., 2007; Meirelles and Raffin, 2017; Massaro et al., 2018; García-Villén et al., 2019). One of us has already experimentally studied the potential effect of using montmorillonite as an excipient and found that it increases the praziquantel solubility (Borrego-Sánchez et al., 2017; Borrego-Sánchez et al., 2018; Borrego-Sánchez et al., 2020a).

Currently, knowledge on the interaction of organic molecules, such as praziquantel, with excipients that have complex structures like pores, nanotubes or solid interlayers is still being built up. The investigation of the mechanism of drug release would profit from the use of computational chemistry techniques able to investigate the drug-clay complexes. One of us has already performed static calculations using classical or quantum mechanical approaches (Borrego-Sánchez and Sainz-Díaz, 2021). In addition, investigations of the local dynamical interactions have been reported (Borrego-Sánchez and Sainz-Díaz, 2021). However, these latter studies were limited by the relatively small timescale explored. Since the drug release takes place on a time that exceeds what is possible, it is necessary to go beyond standard molecular dynamics (MD) simulations.

To calculate the rates, we apply and compare the performance of two enhanced sampling methods that are designed to overcome the time scale barrier. One is the Gaussian Mixture-Based Enhanced Sampling (GAMBES) (Debnath and Parrinello, 2020). The other is a variant of the On-the-fly Probability Enhanced Sampling (OPES) method, namely OPES flooding (OPES$_f$) (Invernizzi and Parrinello, 2020).

Through the application of these enhanced sampling simulations to the release of praziquantel from the surface and the interlayer space of montmorillonite, we aim at indicating a viable computational strategy to be applied in other drug release simulations.

## 2. Methodology

We studied two different systems, in one the drug is absorbed on the top surface of a montmorillonite, in the other the drug sits in between the layers. In the first case, the dynamics are fast, and the release can be simulated with standard molecular dynamics. In the second case, the release time is too long to be simulated in this way, thus it required the use of enhanced sampling methods. Here, we summarize the methods used in this second case.

Over the years, many enhanced sampling methods have been developed in the last decades for the study of rare event processes (see, for instance, references: Torrie and Valleau, 1977; Mezei et al., 1997; Laio and Parrinello, 2002; Marsili et al., 2006; Babin et al., 2008; Barducci et al., 2008; Maragakis et al., 2009; Valsson and Parrinello, 2014; Valsson et al., 2016; Fort et al., 2017; Valsson and Parrinello, 2020). However, most of these methods were designed for calculating static properties like free energy differences and these methods and others like parallel tempering alter the dynamics.

Luckily, some of these methods can be engineered so as to be able to compute the rates. This was made possible by the observation made by Grubmüller (1995) and Voter (1997) that dynamic properties can still be extracted from biased simulations that are accelerated by the addition of an external bias $V(\boldsymbol{R})$ function of the atomic coordinates $\boldsymbol{R}$ provided that such bias is null the transition state region. In such a case, a simple relation holds between the physical time $\tau$, the apparent escape time $\tau_{MD}$ and the bias deposited $V(\boldsymbol{R})$:

$$\tau = \left\langle e^{\beta V(\mathbf{R})} \right\rangle_V \tau_{MD} \tag{1}$$

where the average is over the bias simulation and $\beta = (k_B T)^{-1}$ is the inverse Boltzmann factor.

Here we shall use enhanced sampling methods, in which the bias dependence on $\boldsymbol{R}$ is mediated via a set of functions of $\boldsymbol{R}$. Depending on the method, these functions are referred to as descriptors $\boldsymbol{d}(\boldsymbol{R})$ or as collective variables $\boldsymbol{s}(\boldsymbol{R})$. For this type of bias, many suggestions have been made to design a potential that satisfies the Grubmüller and Voter conditions (Tiwary and Parrinello, 2013; Salvalaglio et al., 2014; McCarty et al., 2015). In this work, we use and compare two such enhanced sampling methods, Gaussian Mixture Based Enhanced Sampling (GAMBES) (Debnath and Parrinello, 2020) and On-the-fly Probability Enhanced Sampling flooding approach (OPES$_f$) (Invernizzi and Parrinello, 2020), which ensure in a relatively simple way that no bias is added to the transition state. We outline here these two methods that differ on the way the bias is constructed. The description of the two methods simplifies considerably if we limit ourselves to describe how to use them to compute the escape times from the bound state and not to reconstruct the full free energy landscape. We refer to the original literature for a full description of the two methods.

*GAMBES*

In GAMBES, one introduces $N_d$ descriptors $\boldsymbol{d}(\boldsymbol{R}) \equiv \{d_p(\boldsymbol{R}), p = 1,2 \dots N_d\}$ able to characterize the initial state and performs an unbiased simulation in the bound state. From these data, the probability density $p(\boldsymbol{d}(\boldsymbol{R}))$ is estimated by fitting the data to a Gaussian mixture (Schwarz, 1978). A bias is then constructed using the relation $V(\boldsymbol{d}) = \left(\frac{1}{\beta}\right) \log(P(\boldsymbol{d}) + \epsilon)$ where $\epsilon$ is a smoothing parameter that also limits the amount of bias that is added. This is very helpful in making so that the conditions for the validity of Eq. 1 are satisfied.

*OPES*

The OPES method is also based on building a bias starting from an estimate of the probability distribution (Invernizzi and Parrinello, 2020). Rather than using descriptors that can be very many, one uses like in umbrella sampling or metadynamics a small number of collective variables $\boldsymbol{s} = \boldsymbol{s}(\boldsymbol{R})$. However, in the spirit of metadynamics the probability $\mathrm{P}(\boldsymbol{s})$ is constructed on the fly as a linear combination of multi variate Gaussians. The bias is constructed in such a way as to a modify in a preassigned way ($\mathbf{P}^{tg}(\boldsymbol{s})$) the $\boldsymbol{s}$ probability distribution. We shall refer to $\mathrm{P}^{tg}(\boldsymbol{s})$ as the target distribution. Here, we shall use as target distribution the so called well-tempered one $P(s) \propto [\mathrm{P}(\boldsymbol{s})]^{\frac{1}{\gamma}}$ where the parameter $\gamma > 1$ regulates the broadening of the target distribution.

The estimate of the probability is periodically updated and at iteration $n$ is written as $\mathrm{P}_n(\boldsymbol{s}) = \frac{\sum_k^n w^k G(\boldsymbol{s},\boldsymbol{s_k})}{\sum_k^n w^k}$, where $G(\boldsymbol{s}, \boldsymbol{s_k})$ are multivariate Gaussian kernels centered at the value assumed by the collective variable at the previous steps $k$, the weights $w_k = e^{\beta V_{k-1}(\boldsymbol{s_k})}$ are computed from the bias at step $k - 1$, and bias is updated as $V_n(\boldsymbol{s}) = \left(1 - \frac{1}{\gamma}\right)\frac{1}{\beta}\log\left(\frac{\mathrm{P}_n(\boldsymbol{s})}{Z_n} + \epsilon\right)$. In the bias expression $\epsilon$ is a smoothing parameter like the one used in GAMBES and $Z_n$ a normalizing factor.

To calculate the rates, we use the OPES flooding variant (OPES$_f$) that is a modification of OPES designed to avoid depositing bias in the transition region (McCarty et al., 2015; Invernizzi and Parrinello, 2020). As in GAMBES, the $\epsilon$ is used not to overflow the basin. Furthermore, the parameter *EXCLUDED_REGION* can be used to prevent OPES$_f$ from depositing bias in the preassigned region of the configurational space corresponding to the transition state.

In both cases, several calculations are started in the bound basin and the statistical distribution of the exit times $\tau$ are analyzed using the Kolmogorov-Smirnov test to ensure that is Poissonian as appropriate for a rare event scenario.

### 2.1. The Model

#### 2.1.1. The surface model system

The montmorillonite surface was modeled as a slab containing 6x4 layer unit cells to which periodic boundary conditions were applied in the x,y plane. The resulting periodically repeated unit in the supercell had stoichiometry $Na_{24}(Al_{76}Mg_{20})(Si_{188}Al_4)O_{480}(OH)_{96}$. The drug was positioned on the layer surface and it was immersed in a bath of 1300 water molecules. The periodicity along the direction $z$ perpendicular to the surface was 45.90 Å (Figure 1).

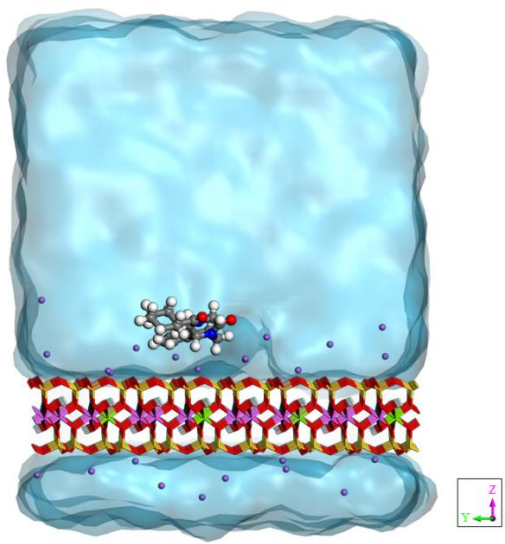

**Figure 1.** Model of praziquantel adsorbed on the montmorillonite surface in aqueous solution.

#### 2.1.2. Model system for the case of the interlayer adsorbed drug

We simulated a molecule of praziquantel adsorbed in a space between two montmorillonite layers immersed in water (Figure 2). In particular, a 6x4x2 supercell of montmorillonite was created with the same composition of a previous experimental work (Borrego-Sánchez et al., 2018). The edges of both layers were cleaved along the (010) and (01̄0) planes to break the periodicity of the layers along the $y$ direction. The valence of the oxygen atoms was completed by adding hydrogen atoms. During the simulations, the structural integrity of the clay edges was preserved. The terminating hydrogens were assigned a charge of +0.338 to neutralize the total charge of the clay, also taking into account that the negative structural charge of the system. Therefore, the chemical formula of the resulting montmorillonite crystal is $Na_{48}(Al_{152}Mg_{40})(Si_{376}Al_8)O_{900}(OH)_{312} \cdot 96\ H_2O$. Both layers are identical and each interlayer space has 48 waters, that is, 2 waters per sodium according to experiments (Borrego-Sánchez et al., 2018). In one of the interlayer spaces, the praziquantel molecule was adsorbed and 2881 water molecules were placed outside the clay as shown in Figure 2. Periodic boundary conditions were applied. The box size was $L_x$=30.96, $L_y$=128.06, $L_z$=30.00, $\alpha$=90.00, $\beta$=100.46 and $\gamma$=90.00 (distances in Å and angles in °).

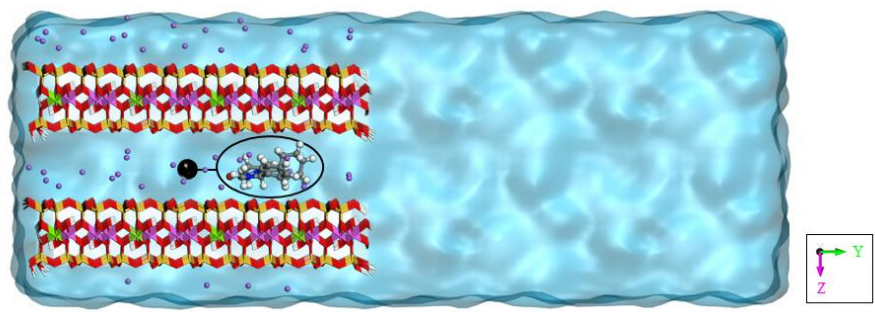

**Figure 2.** Model of praziquantel adsorbed in the interlayer space of montmorillonite in aqueous solution. Here, only part of the waters are shown. The geometrical descriptor used in the GAMBES and OPES$_f$ simulations is highlighted, corresponding to the $y$-component of the vector connecting the center of mass of the drug molecule and a fixed point in the middle of the clay interlayer space ($X^0$) (see text).

Setting up the system for the study of the drug release from the interlayer region was difficult due to the small system size and the limitation of the force field. The main problem was that, when we set up the simulations, some of the counterions left the interfacial region and swelling started, being the counterions essential for holding together the layers. To avoid this unwanted effect, we limited the interlayer distances. However, despite this precaution, a relative sliding of the two layers was observed. Since we deemed this to be an effect of the system's finite size, unlikely to take place in macroscopic systems. This also forced us to fix this degree of freedom. An account of the attempt made can be found in the Supplementary Information.

### 2.3. Computational details

The simulations were driven by the LAMMPS (Plimpton, 1995) suit of programs interfaced with the metadynamics plugin PLUMED (Tribello et al., 2014). The force field used was the consistent-valence force field, also called cvff interface (CVFF) (Heinz et al., 2006; Heinz et al., 2013) that describes the interaction of layered phyllosilicates with organic compounds. The atomic charges of the montmorillonite were set as in Heinz and Suter (2004). This set up has been used elsewhere to describe the clay structure and that of organic molecules (Borrego-Sánchez et al., 2016; Borrego-Sánchez et al., 2020b). All simulations were performed at the physiological temperature of 310 K.

The equilibration of the system with the drug adsorbed on the clay's surface consisted of 10 ps NPT dynamic simulations and followed by other 10 ps in the NVT. Subsequently, to determine the release time, we collected statistics from 10 unbiased simulations of 1 ns long using the orthorhombic version of the Parrinello-Raman barostat (Parrinello and Rahman, 1981).

In the case of the intercalated drug, after a complex equilibration to stabilize the system (see Supplementary Information), we fixed the layers of the clay (zero forces and zero velocities) to prevent their movement. The value used for the basal spacing $d(001)$ was 16 Å. It corresponds to the $d(001)$ measured experimentally for the praziquantel-montmorillonite systems (Borrego-Sánchez et al., 2018; Borrego-Sánchez et al., 2020a). Subsequently, to calculate the drug release kinetics, we ran 25 independent biased NVT simulations up to 100 ns long.

In GAMBES, we used only one descriptor $d$ that is the $y$-component of the vector connecting the center of mass of the drug molecule and a fixed point in the middle of the clay interlayer space ($X^0$). The biased simulations started with the drug at $X^0$, from which it diffuses before escaping. The static bias $V(d)$ was constructed to act only on this known state and to drive the drug release process. To limit the bias deposition, the energy cutoff related to the $\epsilon$ parameter was 7 kcal/mol. This value allowed the drug release.

In OPES$_f$, we used as CV $s$ the same variable as in GAMBES. We prevented depositing bias in the region $y > 8.5$ or $< -8.5$ Å (*EXCLUDED_REGION*). To calculate $\tau$ of different structures and then the diffusion coefficient, the cutoff was 7 kcal/mol when the starting point of the molecule was at $X^0$ (Figure 3A) and also when it was between $X^0$ and the edge of the clay (Figure 3B). No bias was required when it was in the edge of the interlayer space (Figure 3C).

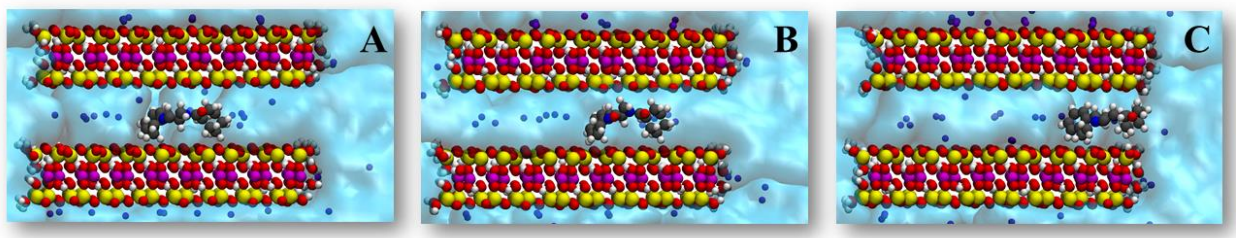

**Figure 3.** Different initial structures with the praziquantel placed in the center (A), between the center and the edge (B), and in the edge (C) of the montmorillonite interlayer space.

In the Supplementary Information, we show that the GAMBES and OPES$_f$ methods give very close results. However, in this application, OPES$_f$ appears to be more efficient. The results presented in the main text are based on OPES$_f$.

## 3. Results

### 3.1. Drug adsorbed on the clay's surface

We performed 10 unbiased simulations of praziquantel adsorbed on the montmorillonite surface. The drug showed a fast desorption from the clay into the water with a computed release time of 363 ps (see Table 1).

**Table 1.** Drug release time ($\tau$) and rate ($k = 1/\tau$). p-value measures the quality of the fit using the Kolmogorov-Smirnov analysis. We also present the average release time $\mu$ and its standard deviation $\sigma$.

|  | $\tau$ ($10^{-12}$ s) | $k$ ($10^9$ s$^{-1}$) | p-value | $\mu \pm \sigma$ ($10^{-12}$ s) |
|---|---|---|---|---|
| **Surface, MD** | 363 | 2.76 | 0.76 | $344 \pm 218$ |

Selected snapshots from a representative release trajectory are displayed in Figure 4. In the initial structure (Figure 4, panel 1), the drug is adsorbed on the surface of the clay in an orientation almost parallel to the clay surface and a planar conformation. Subsequently, the drug adopts a perpendicular orientation (Figure 4, panel 2) and a bent conformation. Finally, the praziquantel completely loses its interaction with the montmorillonite surface and the drug is desorbed (Figure 4, panel 3). In ~71 % of the cases, the aromatic ring is the last to be released. In the other cases, instead, the aliphatic part is released last.

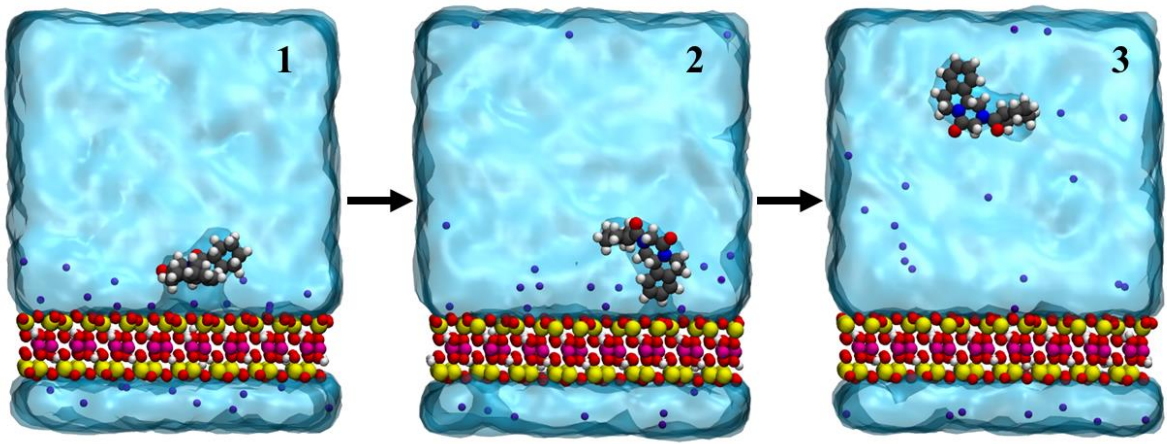

**Figure 4.** Desorption of praziquantel from montmorillonite surface in aqueous solution obtained from standard molecular dynamics simulations.

These results indicate that in the physiological environment the praziquantel molecules will be immediately released.

### 3.2. Drug adsorbed in the clay's interlayer space

In this case, the drug release process takes place in two steps. In the first, the drug diffuses inside the clay. In the second step, it is released from the edge of the clay to the water. To characterize the kinetics of both steps, the simulations were performed with the drug starting from the three different positions displayed in Figure 3. OPES$_f$ was needed when the molecule was inside the clay and therefore we ran two sets of 25 biased simulations from the structures of Figures 3A and 3B. In the case of the drug positioned at the edge (Figure 3C), observing the release did not require enhanced sampling and we carried out 25 unbiased simulations.

Table 2 shows the time that praziquantel takes to exit to the water solution from the three situations. As can be seen, when it is initially located at the center $X^0$ (Figure 3A), $\tau = 200$ μs. This $\tau$ decreases to 54.4 μs when the molecule starts at a position closer to the edge (Figure 3B). Finally, at the edge, we obtained a $\tau$ value of only 5.47 ns (Figure 3C).

**Table 2.** Drug release time ($\tau$) and rate ($k = 1/\tau$) for structures A, B and C of Figure 3. p-value measures the quality of the fit using the Kolmogorov-Smirnov analysis. We also present the average release time $\mu$ and its standard deviation $\sigma$.

|  | $\tau$ (10$^{-6}$ s) | $k$ (10$^6$ s$^{-1}$) | p-value | $\mu \pm \sigma$ (10$^{-6}$ s) |
|---|---|---|---|---|
| **A, OPES$_f$** | 200.0 | 0.005 | 0.76 | 198.0 ± 176.0 |
| **B, OPES$_f$** | 54.4 | 0.018 | 0.41 | 55.4 ± 64.5 |
| **C, MD** | 0.00547 | 182.8 | 0.42 | 0.00536 ± 0.00393 |

With these results, we observed that diffusion within the clay is the slowest process. To determine the diffusion coefficient $D$ inside the clay, we perform two different calculations that start from two different distances from the edge *y1* and *y2*, where *y2* (Figure 3A) is further away from the rim than *y1* (Figure 3B). If the exit time in these two cases are *t1* and *t2* then we use $D \sim \frac{y^2 - y^1}{t^2 - t^1}$. While not rigorous, this provides a rough estimate of this important parameter. The calculated $D$ was $1 \cdot 10 \cdot 10^{-11}$ cm$^2$ s$^{-1}$. This value is consistent with previous experimental results on a similar organic molecule (tryptophan) trapped in a clay-based material ($D \sim 5 \cdot 10^{-11}$ cm$^2$ s$^{-1}$) (Dutournié et al., 2019). It is five orders of magnitude smaller than the diffusion coefficient of praziquantel dissolved in water (De Jesus et al., 2010; Špehar et al., 2021).

Next, we describe the praziquantel release mechanism. The drug is initially in a parallel orientation interacting with both layers of the clay. Then, it diffuses to the edge always keeping this double binding (Figure 5, panels 1 and 2). Throughout the diffusion process, the oxygen of the carbonyl groups interacts with the silicon atoms of the clay surface. In addition, the carbonyl group negative charges are screened by sodium cations and water. However, before exiting, the drug interacts with only one layer (Figure 5, panel 3). Water molecules solvate it, favoring its final release (Figure 5, panel 4). Most of the time the aromatic ring is most frequently the last to be released into the solution. This behavior is similar to that observed at the surface. Once the praziquantel is released, the cations cease to help screening the carbonyl charges, a task that is from now on left to the water molecules.

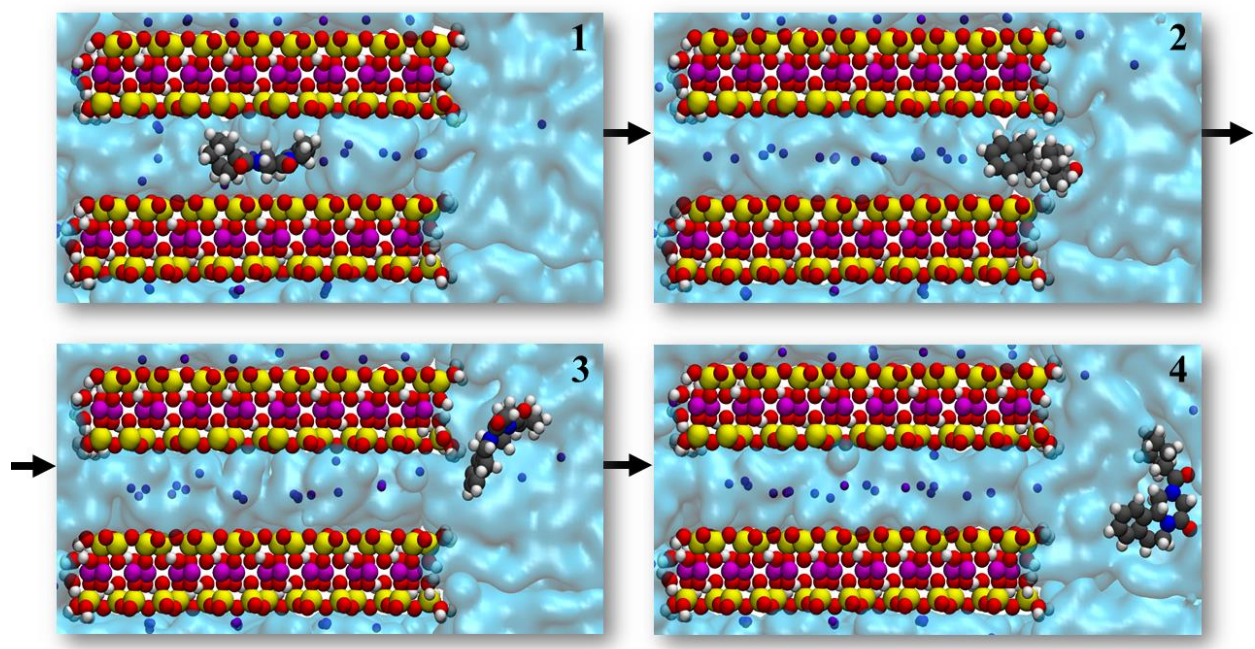

**Figure 5.** Praziquantel release mechanism from the interlayer space of the montmorillonite in aqueous solution.

## 4. Conclusions

This paper shows that despite the shortcoming of the potentials, valuable information can be obtained from molecular dynamics calculations. Our main finding is that in the case in which the praziquantel molecule is inserted in the interlayer regions, the rate limiting step is the drug diffusion toward the water solution. Once the drug is at the layer edge the drug release is extremely fast, of the order of a few hundredth picoseconds. Equally fast is the desorption from the external clay surface.

This suggests several strategies to modulate the release time. For instance, one could search for ways of controlling the penetration length inside the clay. Attempts could also be made at regulating the interlayer distance by means of appropriate spacers (Lagaly et al., 2006; Zhang et al., 2008; He et al., 2013; Chiu et al., 2014) or by using other clays with different interlayer spacing (Odom, 1984; Kevan, 2003; Aguzzi et al., 2007; Bleam, 2012). In the latter two cases, our approach will allow different candidates to be screened before performing the experiments.

## Funding

This project has received funding from the European Union's Horizon 2020 research and innovation programme under the Marie Sklodowska-Curie grant agreement No 895355.

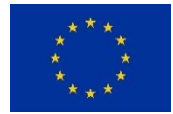

## Acknowledgements

We gratefully acknowledge the HPC infrastructure and the Support Team at Fondazione Istituto Italiano di Tecnologia. ABS would like to acknowledge the discussions and helpful suggestions from Valerio Rizzi and Nicola Tirelli.